\documentclass{aa}  
\usepackage{graphicx}
\usepackage{txfonts}
\newcommand{\av}{$A_V$}

\newcommand{\lsun}{L$_{\odot}$}

\newcommand{\msun}{M$_{\odot}$}
\newcommand{\rsun}{R$_{\odot}$}
\newcommand{\msunyr}{M$_{\odot}$\,yr$^{-1}$}
\newcommand{\macc}{$\dot{M}_{acc}$}

\newcommand{\lacc}{$L_{\mathrm{acc}}$}

\newcommand{\rstar}{$R_{\mathrm{*}}$}

\newcommand{\mstar}{$M_{\mathrm{*}}$}

\newcommand{\oi}{\ion{O}{i}}

\newcommand{\fei}{\ion{Fe}{i}}

\newcommand{\hi}{\ion{H}{i}}
\newcommand{\hei}{\ion{He}{i}}

\begin{document}

\title{Do subsequent outbursts of the same EXor source present similar features?}
\author{T. Giannini\inst{1},  A. Giunta\inst{2}, D. Lorenzetti\inst{1}, G. Altavilla\inst{1,2}, S. Antoniucci\inst{1},  F. Strafella\inst{3},  V. Testa\inst{1}
}
\institute{INAF-Osservatorio Astronomico di Roma, via Frascati 33, I-00078 Monte Porzio Catone, Italy \\
              \email{teresa.giannini@inaf.it}         
\and
ASI - Agenzia Spaziale Italiana - Via del Politecnico, 00133 Roma, Italy
\and
Dipartimento di Matematica e Fisica, Universit{\'a} del Salento, 73100 Lecce, Italy
}
\date{Received date / Accepted date}
\titlerunning{V1118 Ori : 2019 outburst}
\authorrunning{Giannini et al.}

\abstract
{V1118 Ori is a classical EXor source whose light curve has been monitored over the past thirty years (although not continuously). It underwent a powerful outburst in 2005, followed by ten years of quiescence and a less intense outburst in 2015. In 2019, a new intense brightness increase was observed  ($\Delta g$ $\sim$ 3 mag).}
{This new accretion episode offers the opportunity to compare the photometric and spectroscopic properties of different outbursts of the same source. This allows one to highlight differences and similarities among different events by removing any possible bias related to the intrinsic properties of the star-disk system.}   
{We discovered the 2019 V1118 Ori outburst by examining the $g$-band light curve acquired by the \textit{Zwicky Transient Facility} and followed the decreasing phase with the \textit{Rapid Eye Mount} telescope in the $griz$ bands. Two near-infrared spectra were also acquired at different brightness stages with the \textit{ Large Binocular Telescope}.}
{The last event shows the following characteristics: 1) amplitude similar than in 2015 and lower than in 2005; 2) duration less than one year as in previous events; 3) rise (decline) speed of 0.018 (0.031) mag/day, which is different from previous cases; 4) a gradual blueing of the [$g-r$] color is observed over time, while the [$r-i$] color remains roughly unchanged; 5) with few exceptions, the near-infrared lines (mainly \hi\, recombination) are the same observed in 2015; 6) the mass accretion rate peaks at \macc\, $\sim$ 10$^{-7}$ \msunyr\, and decreases in about a month down to a few 10$^{-8}$ \msunyr.}
{Our analysis shows that the comparison of data from different outbursts of the same source is a non-trivial exercise, which allows obtaining important clues useful to drive theoretical efforts towards a better  understanding of the EXor phenomenon.}

\maketitle

\section{Introduction}\label{sec:sec1}

Eruptive variables belong to a class of Pre-Main Sequence objects usually divided 
into two sub-classes: EXor (Herbig 1989) and more energetic FUor (Hartmann \& Kenyon 1985) systems that both present an 
intermittent variability related to an unsteady mass accretion rate. The membership 
to one or the other sub-class depends on the different properties, such as bursts 
amplitude and duration, recurrence time between subsequent bursts, mass accretion 
rate. A comprehensive account for the properties of both EXors and FUors is given by 
Audard et al. (2014).

In general terms, the observed features that define the EXor-class membership are: \textit{ (i)} occurrence on time scales of years of short-term outbursts (typically lasting several months) with amplitude $\Delta V$ of about 3-5 mag at optical and near-IR wavelengths; \textit{ (ii)} spectra characterized by strong \hi\, recombination lines and other atomic permitted lines (e.g. Herbig 2008, Lorenzetti et al. 2009, Sicilia-Aguilar et al. 2012, K\'{o}sp\'{a}l et al. 2011). 

Although EXor outbursts are known to be associated with magnetospheric accretion events from the circumstellar disk (Shu et al. 1994), their nature is still very uncertain and no specific models have been developed so far. Therefore, the theoretical framework developed by Hartmann \& Kenyon (1985) for FU Orionis objects is widely adopted also for EXors. More specific efforts have been performed in the last decade (D'Angelo \& Spruit 2010; Vorobyov \& Basu 2015), however, difficulties arise in predicting the relatively short cycle time of the episodic accretion phenomena and there are only suggestions about the mechanism responsible for the outburst onset: thermal instabilities (e.g. Bell \& Lin 1994), presence of a massive planet (Lodato \& Clarke 2004), and close interactions in a binary system (Bonnell \& Bastien 1992). 

Observational inputs to the current theoretical framework can be obtained by comparing the results of both photometric and spectroscopic observations at optical and near-IR frequencies of successive outbursts of the same source. This comparison allows us to highlight similarities and differences of different events
in the same star-disk system  and to remove all possible biases often introduced when considering outbursts of sources with different  mass, luminosity, or evolutionary stage. In this way, the uncontaminated interplay between the outburst parameters (amplitude, duration of both increasing and fading phases, fluctuations of the local extinction) can be more accurately determined.

A suitable target for this kind of investigation is the classical EXor V1118 Ori, a source that has been monitored over the past thirty years by several authors (Audard et al. 2005, 2010 and references therein).
We have been following V1118 Ori for more than ten years observing both quiescent and outbursting periods (Lorenzetti et al. 2006, 2007, 2015). In particular, our monitoring of the outburst occurred in the period 2015-2016 allowed us to estimate the evolution of the mass accretion rate  (compared to the immediately precedent quiescence) from 0.3-20 10$^{-8}$ to 0.2-2 10$^{-6}$ M$_{\sun}$ yr$^{-1}$ (Giannini et al. 2016, 2017). During 2017-2018 V1118 Ori went through a period of quiescence. In May 2019 we announced a new outburst (Giunta et al. 2019a), remarking its peculiarity in terms of rising and declining speed (Giunta et al. 2019b). Our observations of this last event are presented in the following Sect.2, while the results are discussed and compared with those of the previous outbursts in Sect.3. Our final remarks are given in Sect.4.

\section{Observations and data reduction}\label{sec:sec2}

\subsection{Optical Photometry}\label{sec:sec2.1}
In the period 2018-2019 V1118 Ori was monitored in the $g$-band with the \textit{ Zwicky Transient Facility} (ZTF)\footnote{https://www.ztf.caltech.edu}, which reported an increase of about 2 magnitudes (from $g$ $\sim$ 17 to 15) between February and April 2019 (see Table\,\ref{tab:tab1}). Due to seasonal observability, subsequent observations could only start in late August, when the source was still bright ($g$ $\sim$ 14.4), but already in the declining phase. The quiescence was reached  again in late November.

To increase the temporal coverage, we observed V1118 Ori with the ROSS2 optical camera installed on the  60\,cm robotic telescope \textit{ Rapid Eye Mount} (REM) at La Silla (Chile). The observations were conducted in the  $g,r,z,i$ bands, and cover mainly the declining phase from August to September 2019. 

The images were reduced by using standard procedures for bad pixel removal, flat fielding, and bias and dark subtraction. The aperture photometry was then obtained using as calibrators several stars in the V1118 Ori field, whose optical magnitudes have been retrieved by the PAN-STARSS Data Release 2 (DR2) Photometric Catalog \footnote{https://catalogs.mast.stsci.edu/panstarrs/}. The photometric data are listed in Table\,\ref{tab:tab2}. 

\begin{figure}
   \centering
   \includegraphics[width=10cm]{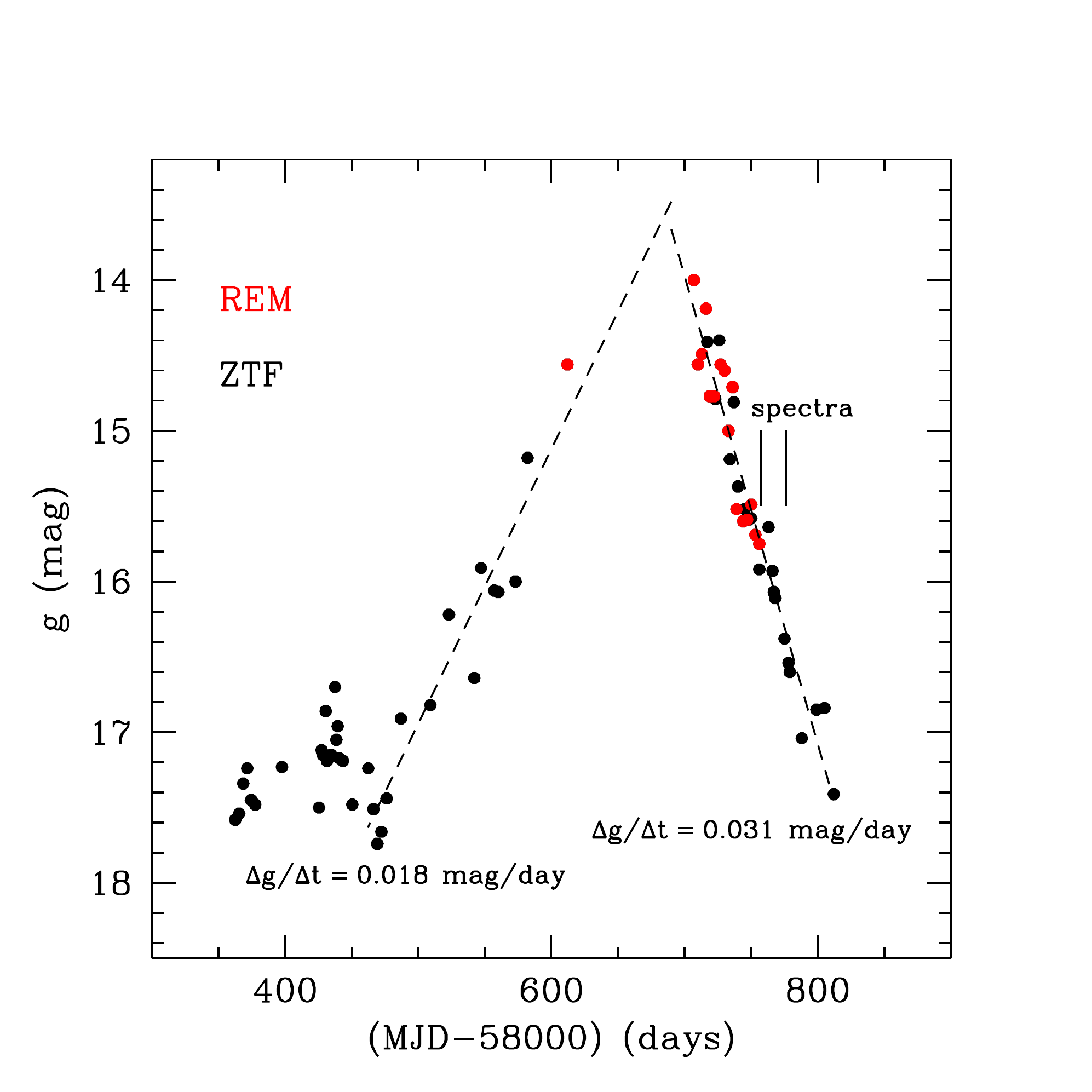}
   \caption{ \label{fig:fig1} ZTF (black) and REM (red) $g$-band photometric points of the 2019 outburst of V1118 Ori. The two dashed lines are the linear fits to the points of the rising and declining phase, with indicated the values of their slopes (mag/day). Both have an uncertainty of 0.002 mag/day. Two continuous segments indicate the dates of the spectroscopic observations.}
 \end{figure}

In Figure~\ref{fig:fig1} the $g$-band light curve is depicted. We have certainly missed the outburst peak, which  occurred in the period June-August 2019. Unfortunately, no \textit{ Gaia} passages on V1118 Ori were performed during this period, being the closest in time carried out on August 13 (Lattanzi, priv. commun.). No alert was issued at that time. The ROSS2 observations have partially filled the observational gap, allowing us to get the highest photometric point of both the rising and declining phases (9 May 2019, $g$\,=\,14.56, and 12 August 2019, $g$\,=\,14.00).

\begin{figure*}
   \centering
   \includegraphics[width=18cm]{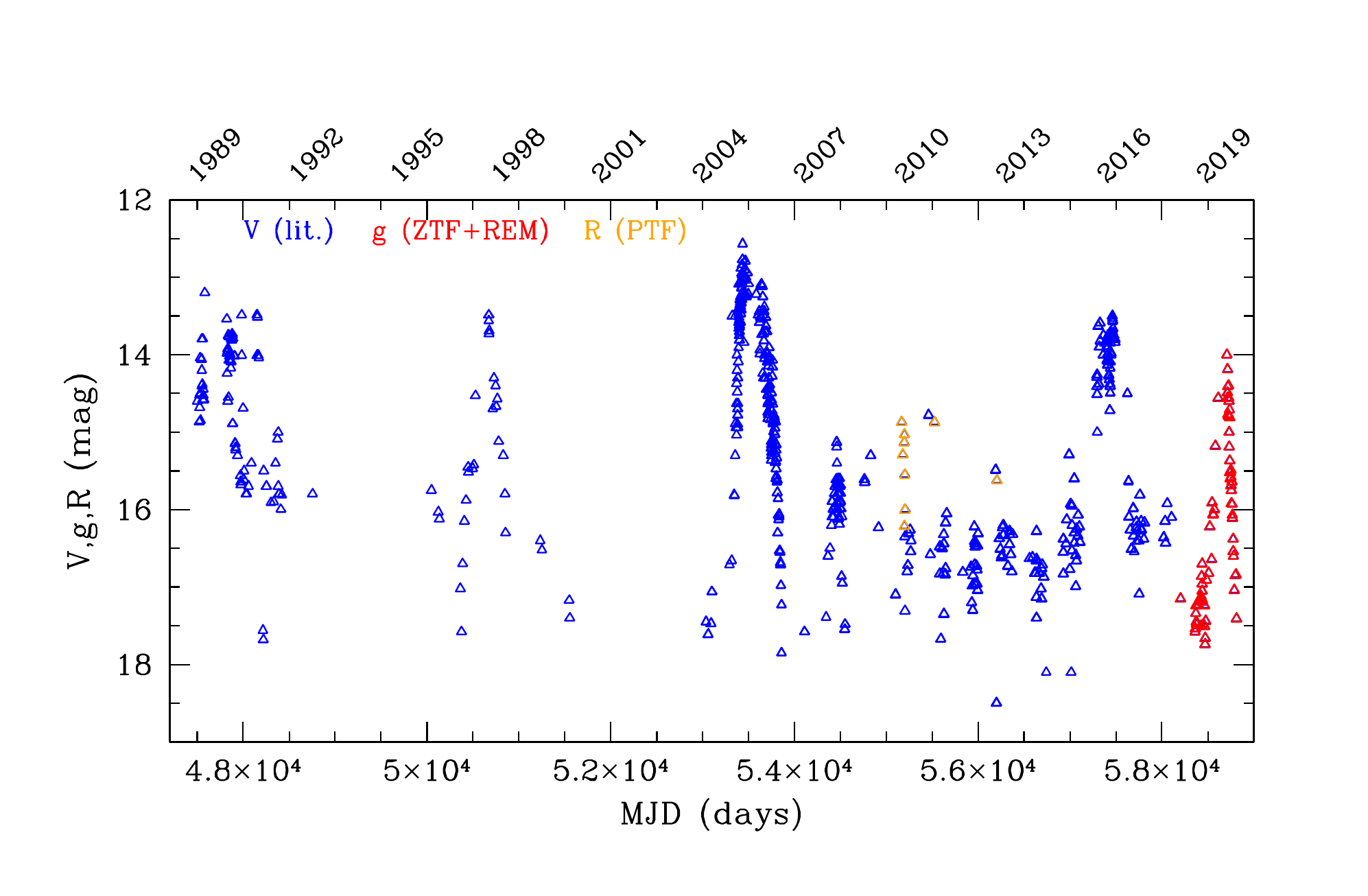}
   \caption{ \label{fig:fig2} Last 30-years optical light curve of V1118 Ori. $V$-band data are taken from Giannini et al.(2017) and references therein (mainly Audard et al. 2010), while in orange are depicted $R$-band photometric points retrived from the \textit{ Palomar Transient Factory} survey. In red are depicted the  ZTF+REM $g$-band data of the 2019 outburst.}
 \end{figure*}

To get a rough estimate of the peak brightness, we have fitted with a straight line both the rising and declining light curves, which intersect at $g$ $\sim$ 13.8 mag (see Sect..\ref{sec:sec3.1.1}). This value is comparable to the peak reached in the 2015-2016 outburst (Giannini et al. 2016), as shown also in the historical light curve presented in Figure\,\ref{fig:fig2}.

%-------------------------------------- Figura 1

%-------------------------------------- Figura 2

\begin{table}
\caption{\label{tab:tab1}ZTF photometry.}
\centering
\begin{tabular}{ccc}
\hline\hline
MJD     &   Date       & $g$       \\
\hline
        &              &  (mag)  \\
\hline\hline        
58204 & 27 Mar 2018  &  17.15  \\
58362 & 01 Sep 2018  &  17.58  \\
58365 & 04 Sep 2018  &  17.54  \\
58368 & 07 Sep 2018  &  17.34  \\
58371 & 10 Sep 2018  &  17.24  \\
58374 & 13 Sep 2018  &  17.45  \\
58377 & 16 Sep 2018  &  17.48  \\
58397 & 06 Oct 2018  &  17.23  \\
58425 & 03 Nov 2018  &  17.50  \\
58427 & 05 Nov 2018  &  17.12  \\
58428 & 06 Nov 2018  &  17.15  \\
58430 & 08 Nov 2018  &  16.86  \\
58431 & 09 Nov 2018  &  17.19  \\
58434 & 12 Nov 2018  &  17.15  \\
58437 & 15 Nov 2018  &  16.70  \\
58438 & 16 Nov 2018  &  17.05  \\
58439 & 17 Nov 2018  &  16.96  \\
58440 & 18 Nov 2018  &  17.17  \\
58443 & 21 Nov 2018  &  17.19  \\
58450 & 28 Nov 2018  &  17.48  \\
58462 & 10 Dec 2018  &  17.24  \\
58466 & 14 Dec 2018  &  17.51  \\
58469 & 17 Dec 2018  &  17.74  \\
58472 & 20 Dec 2018  &  17.66  \\
58476 & 24 Dec 2018  &  17.44  \\
58487 & 04 Jan 2019  &  16.91  \\
58509 & 26 Jan 2019  &  16.82  \\
58523 & 09 Feb 2019  &  16.22  \\
58542 & 28 Feb 2019  &  16.64  \\
58547 & 05 Mar 2019  &  15.91  \\
58557 & 15 Mar 2019  &  16.06  \\
58560 & 18 Mar 2019  &  16.07  \\
58573 & 31 Mar 2019  &  16.00  \\
58582 & 09 Apr 2019  &  15.18  \\
58717 & 22 Aug 2019  &  14.41  \\
58723 & 28 Aug 2019  &  14.79  \\
58726 & 31 Aug 2019  &  14.40  \\ 
58734 & 08 Sep 2019  &  15.19  \\
58737 & 11 Sep 2019  &  14.81  \\
58740 & 14 Sep 2019  &  15.37  \\
58745 & 19 Sep 2019  &  15.52  \\
58748 & 24 Sep 2019  &  15.59  \\
58750 & 26 Sep 2019  &  15.58  \\
58756 & 02 Oct 2019  &  15.92  \\
58763 & 09 Oct 2019  &  15.64  \\
58766 & 12 Oct 2019  &  15.93  \\
58767 & 13 Oct 2019  &  16.07  \\ 
58768 & 14 Oct 2019  &  16.11  \\
58775 & 19 Oct 2019  &  16.38  \\
58779 & 23 Oct 2019  &  16.54  \\
58780 & 24 Oct 2019  &  16.60  \\
58788 & 01 Nov 2019  &  17.04  \\
58799 & 12 Nov 2019  &  16.85  \\
58805 & 18 Nov 2019  &  16.84  \\
58812 & 25 Nov 2019  &  17.41  \\
\hline\hline
\end{tabular}
\tablefoot{Typical photometric errors are 0.05 mag.}
\end{table}

\begin{table}
\caption{\label{tab:tab2} REM photometry.}
\centering
\begin{tabular}{cccccc}
\hline\hline
MJD     &   Date         & $g$       &  $r$         &  $i$        &   $z$     \\
\hline 
        &                &  (mag)  &  (mag)     &  (mag)    &  (mag)  \\
\hline\hline        
 58612	& 09 May 2019    &  14.56  &    13.83   &   13.31   &  12.76  \\
 58707	& 12 Aug 2019    &  14.00  &    13.59   &   12.99   &  12.60  \\
 58710	& 15 Aug 2019    &  14.56  &    13.81   &   13.45   &  12.87  \\
 58713	& 18 Aug 2019    &  14.49  &    13.74   &   13.34   &  12.74  \\
 58716  & 21 Aug 2019    &  14.19  &    13.57   &   13.05   &  12.64  \\
 58719  & 24 Aug 2019    &  14.77  &    13.99   &   13.48   &  12.96  \\
 58722  & 27 Aug 2019    &  14.77  &    14.01   &   13.47   &  12.97  \\
 58727  & 01 Sep 2019    &  14.56  &    13.73   &   13.32   &  12.83  \\
 58730  & 04 Sep 2019    &  14.60  &    13.85   &   13.42   &  12.91  \\
 58733  & 07 Sep 2019    &  15.00  &    14.32   &   13.86   &  13.20  \\
 58736  & 10 Sep 2019    &  14.71  &    14.02   &   13.48   &  13.01  \\
 58739  & 13 Sep 2019    &  15.52  &    14.62   &   14.24   &  13.52  \\
 58744  & 18 Sep 2019    &  15.60  &    14.79   &   14.27   &  13.52  \\
 58747  & 21 Sep 2019    &  15.59  &    14.74   &   14.41   &  13.72  \\
 58750  & 24 Sep 2019    &  15.49  &    14.82   &   14.33   &  13.71  \\
 58753  & 27 Sep 2019    &  15.69  &    14.99   &   14.49   &  13.85  \\
 58756  & 30 Sep 2019    &  15.75  &    15.06   &   14.62   &  13.88  \\
\hline\hline
\end{tabular}
\tablefoot{Typical photometric errors are 0.05 mag.}
\end{table}

\subsection{Spectroscopy}\label{sec:sec2.2}
Near-IR spectra were collected on two separate occasions (October 1 and October 20, 2019 (respectively MJD 58757 and 58776). Both spectra were obtained with the 8.4\,m \textit{ Large Binocular Telescope} (LBT) located in Mount Graham (Arizona, USA), by using the LUCI2 spectrometer. The observations were conducted with the G200 grating (zJ and HK filters) covering the spectral ranges 0.9$-$1.20 $\mu$m and 1.50$-$2.40 $\mu$m. We used the 0\farcs 75 slit, corresponding to a spectral resolution between 1200 and 1700. The standard ABB$^{\prime}$A$^{\prime}$ technique was adopted to perform the observations, for a total integration time of 16 minutes for each spectrum.

\begin{figure*}
   \centering
   \includegraphics[width=14cm]{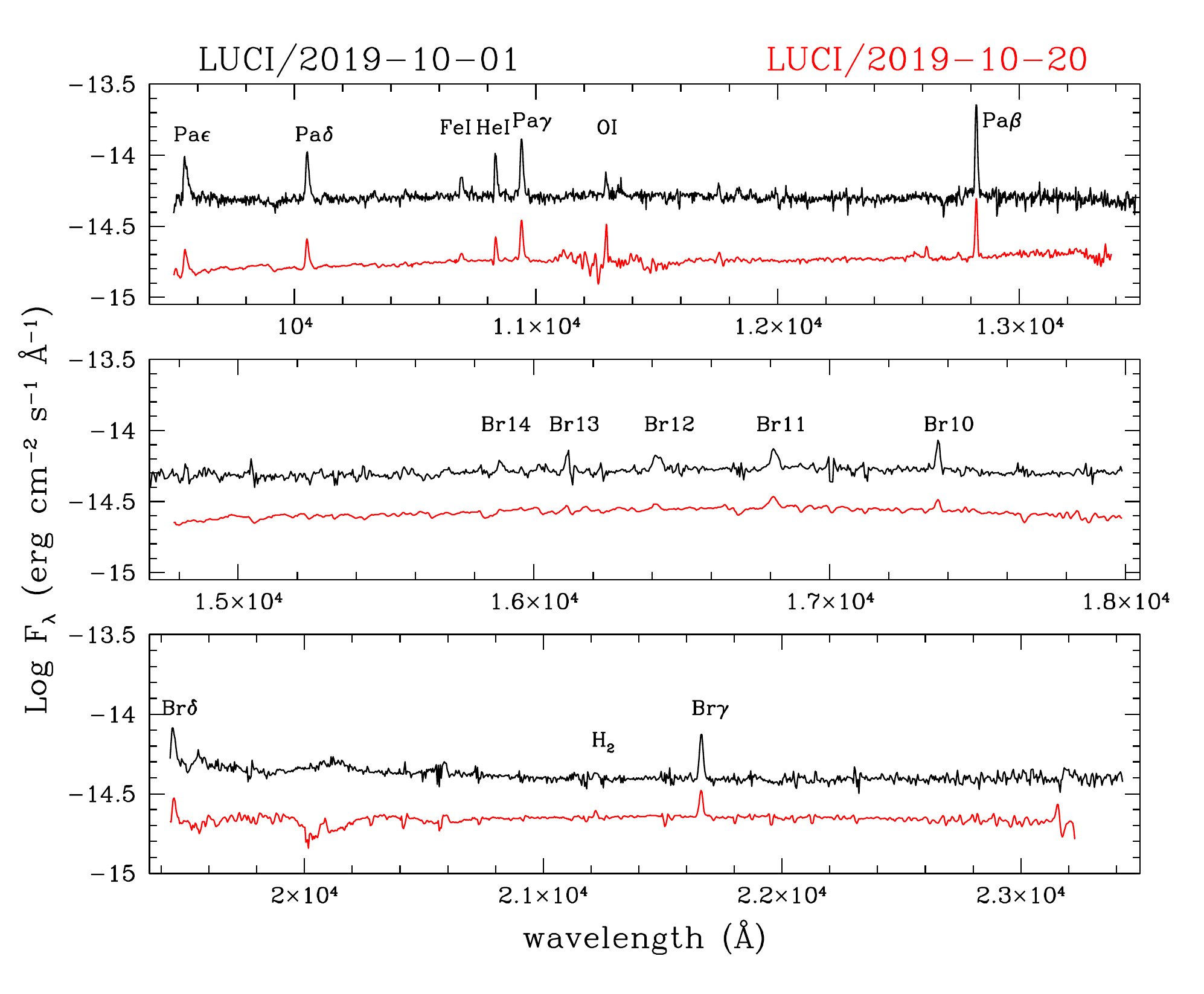}
   \caption{ \label{fig:fig3} Near-IR spectra taken on October 1 (black) and October 20 (red). Main emission lines are labelled.}
 \end{figure*}

Data reduction was carried out at the Italian LBT Spectroscopic Reduction Center\footnote{http://www.iasf-milano.inaf.it/Research/lbt\_rg.html}, using scripts optimized for LBT observations. The spectral images were flat-fielded, sky-subtracted, and corrected for optical distortion in both  spatial and spectral directions. The telluric features were removed by dividing the extracted spectra by that of a telluric standard star (observed on the same nights as the targets), once corrected for its intrinsic hydrogen absorption features. Wavelength calibration was obtained from arc lamps. The $K$-band acquisition image was used to estimate the photometry of V1118 Ori on the dates of the spectral observations ($K$=10.08 on October 1 and $K$=10.70 on October 20), which in turn was used for flux calibration of the HK spectral segments. The zJ spectrum taken on October 1 was calibrated  using the $z$-band photometry of September 30, while the zJ spectrum of October 20 was simply aligned with the HK segment, applying the same calibration factor. 
The resulting 0.95$-$2.35 $\mu$m spectra are shown in Figure~\ref{fig:fig3}, while fluxes of the relevant features identified are listed in Table\,\ref{tab:tab3}.
Since both spectra were acquired during the declining phase, it is not surprising that they are similar to the LUCI2 spectrum obtained during the same phase of the 2015-2016 outburst (Giannini et al. 2017). The most prominent lines are  
the  \hi\, recombination lines of the Paschen and Brackett series, along with a couple of metallic lines of \fei\, and \oi\, which were not detected in 2016. The \hei\, 1.08 $\mu$m line is also  detected in both spectra and exhibits a weak PCyg-like absorption in the October 20 spectrum, which signals outflow activity. The H$_2$ 2.12 $\mu$m emission is detected only when the source is close to quiescence (October 20), as happened in the past (Lorenzetti et al. 2014, Giannini et al. 2017). Likely, the line is not detected during outbursts due to an unfavorable line-to-continuum ratio when the source is bright. The fact that the H$_2$ emission is substantially unrelated with the source brightness, suggests that it comes from the diffuse cloud.

\begin{table}
\caption{\label{tab:tab3} Line fluxes.}
\centering
\begin{tabular}{cccc}
\hline\hline
               &                   &     01 Oct 2019         &  20 Oct 2019      \\
\hline 
 Line          & $\lambda_{vac}$   &    \multicolumn{2}{c}{F$\pm\Delta$ F}\\
               &   (\AA)           &    \multicolumn{2}{c}{(10$^{-14}$ erg s$^{-1}$ cm$^{-2}$)} \\
\hline\hline        
 Pa$\epsilon$  &       9549        &     9.7$\pm$0.6      &   2.19$\pm$0.05      \\
 Pa$\delta$    &      10052        &    10.1$\pm$0.4      &   2.19$\pm$0.05      \\
 \ion{Fe}{i}   &      10691        &     3.4$\pm$0.3      &   0.49$\pm$0.05      \\
 \ion{He}{i}   &      10832        &     6.3$\pm$0.3      & (-0.1)$^a$ 1.00$\pm$0.03 \\
 Pa$\gamma$    &      10941        &    11.7$\pm$0.3      &   3.49$\pm$0.02      \\
 \ion{O}{i}    &      11290        &     2.5$\pm$0.3      &   0.41$\pm$0.03      \\
 Pa$\beta$     &      12821        &    19.7$\pm$0.5      &   3.84$\pm$0.03      \\
 Br14          &      15885        &     2.0$\pm$0.5      &     -                \\
 Br13          &      16114        &     2.5$\pm$0.8      &     -                \\ 
 Br12          &      16413        &     4.5$\pm$1.0      &   1.0$\pm$0.1        \\
 Br11          &      16811        &     5.5$\pm$0.4      &   1.8$\pm$0.2        \\
 Br10          &      17366        &     5.5$\pm$0.4      &   2.0$\pm$0.1        \\
 Br$\delta$    &      19451        &     7.7$\pm$0.6      &   1.7$\pm$0.2        \\
 H$_2$         &      21218        &      $<$0.5$^b$      &   0.3$\pm$0.1        \\
 Br$\gamma$    &      21661        &     6.7$\pm$0.2      &   1.9$\pm$0.1        \\

\hline\hline
\end{tabular}
\tablefoot{ $^a$ PCyg-like absorption; $^b$ 3-$\sigma$ upper limit.}
\end{table}

%-------------------------------------- Figura 3

\section{Results and Discussion}\label{sec:sec3}

\subsection{Photometry}\label{sec:sec3.1}

%-------------------------------------- Figura 4

  \begin{figure}
   \centering
   \includegraphics[width=8cm]{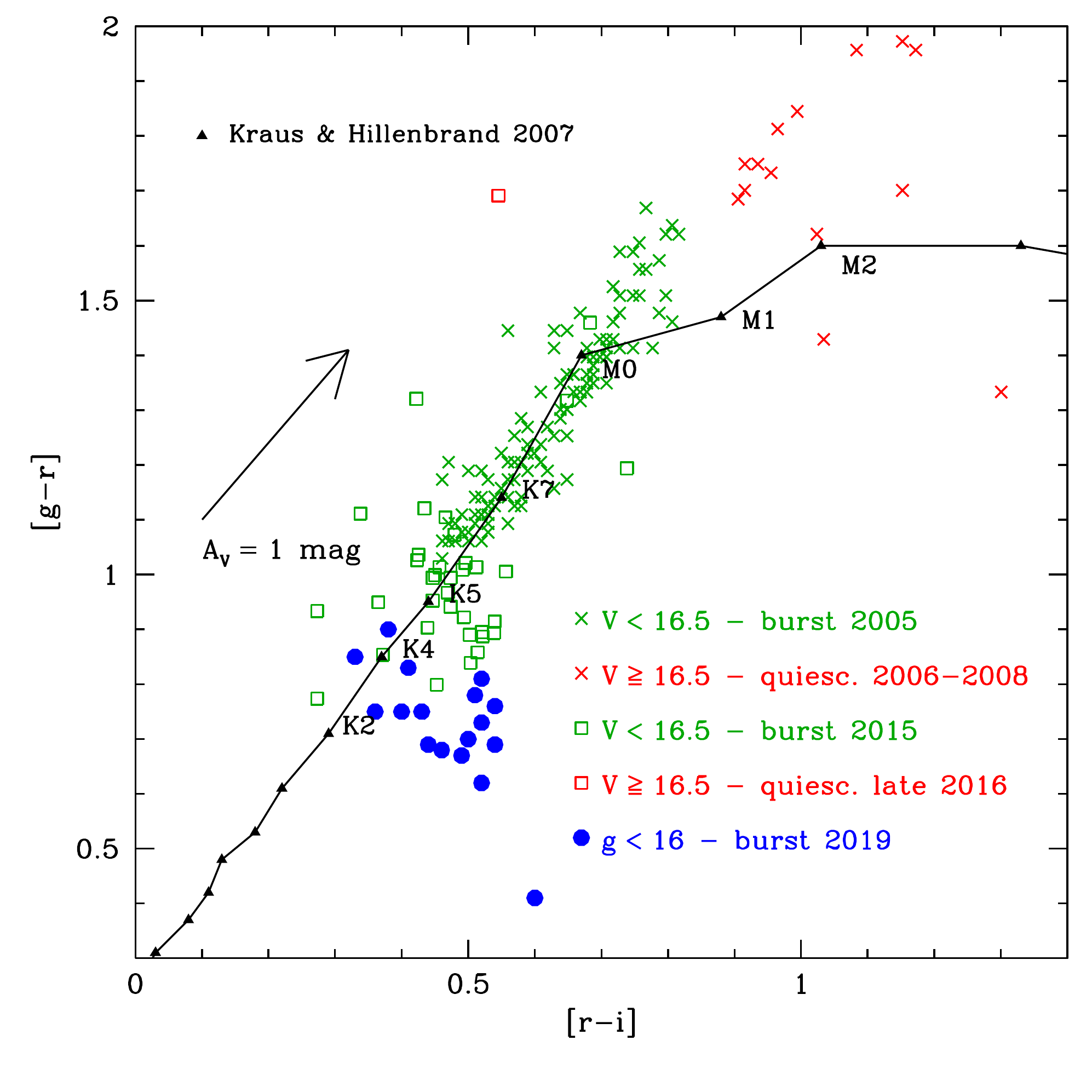}
 \caption{ \label{fig:fig4}$[g-r$] vs. [$r-i$] diagram. Blue dots are the points of the 2019 outburst. In green are the points of the 2005 and 2015 outbursts ($V$ $<$ 16.5, crosses and open squares), while in red are the colors of the subsequent quiescence phases ($V \ge$ 16.5). Transformation equations between Johnson-Cousins and Sloan photometric systems have been applied. The continuos line is the \textit{ locus} of 400-600 Myrs young stars (Kraus \& Hillenbrand 2007). The black arrow is the extinction vector for A$_V$\,=\,1 mag, adopting the Cardelli et al. (1989) extinction law.}
 \end{figure}

\subsubsection{Increasing $vs.$ decreasing speed}\label{sec:sec3.1.1}

%%%%%%%%%%%%%%Table 4 %%%%%%%%%%%%%%%%%%%%%%%%%%%%

\begin{table}
\begin{center} 
\caption{Brightness variation speed observed during the last four outbursts of V1118 Ori. \label{tab:tab4}} 
\medskip
{
\begin{tabular}{cccc}
\hline
\medskip
outburst (year)    & R (rise)      & D (decline)    &  R/D     \\ 
                & \multicolumn{2}{c}{(mag/day)}  &          \\
\hline
    1997$^a$        & 0.011         &  0.013        &  0.85    \\
    2005$^a$        & 0.035         &  0.015        &  2.33    \\
    2015$^a$        & 0.015         &  0.015        &  1.00    \\
    2019$^b$        & 0.018         &  0.031        &  0.58    \\
\hline\hline

\end{tabular}}
\end{center}
%\medskip 
\tablefoot{$^a$ $V$-band data; $^b$ $g$-band data.}
\end{table} 

%%%%%%%%%%%%%%%%%%%%%%%%%%%%%%%%%%%%%%%%%%%%%%%%%%%%%%%%%%%%%%%%%%%%%%%%%%%%%   
As shown in Figure\,\ref{fig:fig2}, well-sampled monitoring of V1118 Ori was obtained for the last five outbursts, from the one occurred in 1997 to that of 2019.
Therefore, V1118 Ori offers the unique possibility of comparing the speed of brightness variation ($\Delta$mag/$\Delta$t) calculated with the same method in subsequent events of the same EXor source.
This speed can be easily obtained as the slope of the linear fit through the photometric points of the increasing and decreasing phase (Figure\,\ref{fig:fig1}).

For the last outburst, we get $\Delta g$/$\Delta$t\,=\,0.018 mag/day (rising) and 0.031 mag/day (declining). 
These have been estimated by considering as starting and ending points of the fit those from which the brightness starts (stops) to increase (decrease) monotonically. Note that this choice implies that the slope of 0.018 mag/day is the maximum speed value that can be fitted through the rising phase data.  

The same method applied to the data of previous outbursts provides the $\Delta V$/$\Delta$t reported in Table\,\ref{tab:tab4}. Broadly speaking, the majority of both increasing and decreasing speed values range between 0.011 and 0.015 mag/day. Two significant exceptions are represented by the increasing speed of the 2005 outburst (0.035 mag/day) and the decreasing speed of the last event (0.031 mag/day). More importantly, the data in Table\,\ref{tab:tab4} show that while in 2005 the increase was faster (by more than a factor two) compared to the decrease, the reverse phenomenon occurred in the last outburst.

These differences provide indications that heating  and cooling occur with different modalities in subsequent outburst events. 
We speculate that a role might be played by the temperature reached at the outburst peak, by the quantity of gas inside the accretion columns and by the fraction of the stellar surface involved in the accretion shocks. Indeed, 
very different scenarios for the outburst dynamics are predicted by theoretical models, from two ordered funnel streams forming two hot spots on the stellar surface (e.g. Romanova et al. 2004) to multiple unstable 'tongues'
of matter that form chaotic spots with irregular shape and position (e.g. Romanova et al. 2008).

\subsubsection{Does an outburst periodicity exists ?}\label{sec:sec3.1.2}

In general, the intermittence of the EXor's outbursts does not show any clear periodicity and, looking at Figure 2, V1118 Ori seems to obey the same trend. However, to have a more quantitative confirmation, we have performed the Lomb-Scargle analysis for all available photometries ($V$ and $g$ bands) collected in the last 30 years. The resulting periodogram is characterized by several peaks with little statistical significance mainly due to a very noisy quiescence level and
unevenly sampled data. In any case, the presence of multiple peaks does not favor a periodical behavior of the outburst events.

\subsubsection{Two-colors plot}\label{sec:sec3.1.3}
The two-color diagram [$g-r$] vs. [$r-i$] derived from REM data (blue dots) is presented in  Figure~\ref{fig:fig4}. For comparative purposes, we also show the colors of the period 2005-2016, taken from Audard et al. (2010) and Giannini et al. (2017). These were obtained from the original Johnson-Cousins photometric points  $VR_cI_c$ transformed into Sloan magnitudes\footnote{http://www.sdss3.org/dr8/algorithms/sdssUBVRITransform.php}.  Different symbols and colors indicate different outburst episodes ($V <$ 16.5) and subsequent post-outburst periods ($V \ge$ 16.5).
Colors of the low-mass young stellar population in Praesepe and Coma Berenices (400-600 Myr) from spectral types F8 to M3 (Kraus \& Hillenbrand 2007) are also depicted as black triangles. The direction of the extinction vector is represented, adopting the Cardelli et al. (1989) extinction law.

The main difference among the three outbursts illustrated in Figure~\ref{fig:fig4} lies in the gradual blueing of the color [$g-r$], which on average is (in mag) 1.29 in 2005, 1.00 in 2015 and 0.73 in 2019. Note that this behavior cannot be attributed to a continuous decrease of the visual extinction, since, as we show in Section\,\ref{sec:sec3.2.2}, the A$_V$ measured during the three outbursts was roughly the same (A$_V$ $\sim$ 1.5 mag). Therefore, we interpret the steadily decreasing [$g-r$] value over the past 15 years with a gradual increase in the temperature of the photosphere during subsequent outbursts. 
Vice versa, the color
[$r-i$] remains almost unchanged during the three outbursts (the average [$r-i$] values are 0.62, 0.47, and 0.47 mag in 2005, 2015, and 2019, respectively). 
Therefore, the variations of [$r-i$] are likely to be driven by mechanisms not closely related to the outburst episodes.

\subsection{Near-IR Spectroscopy}\label{sec:sec3.2}
%-------------------------------------- Figura 5
\begin{figure}
\centering
\includegraphics[width=9cm]{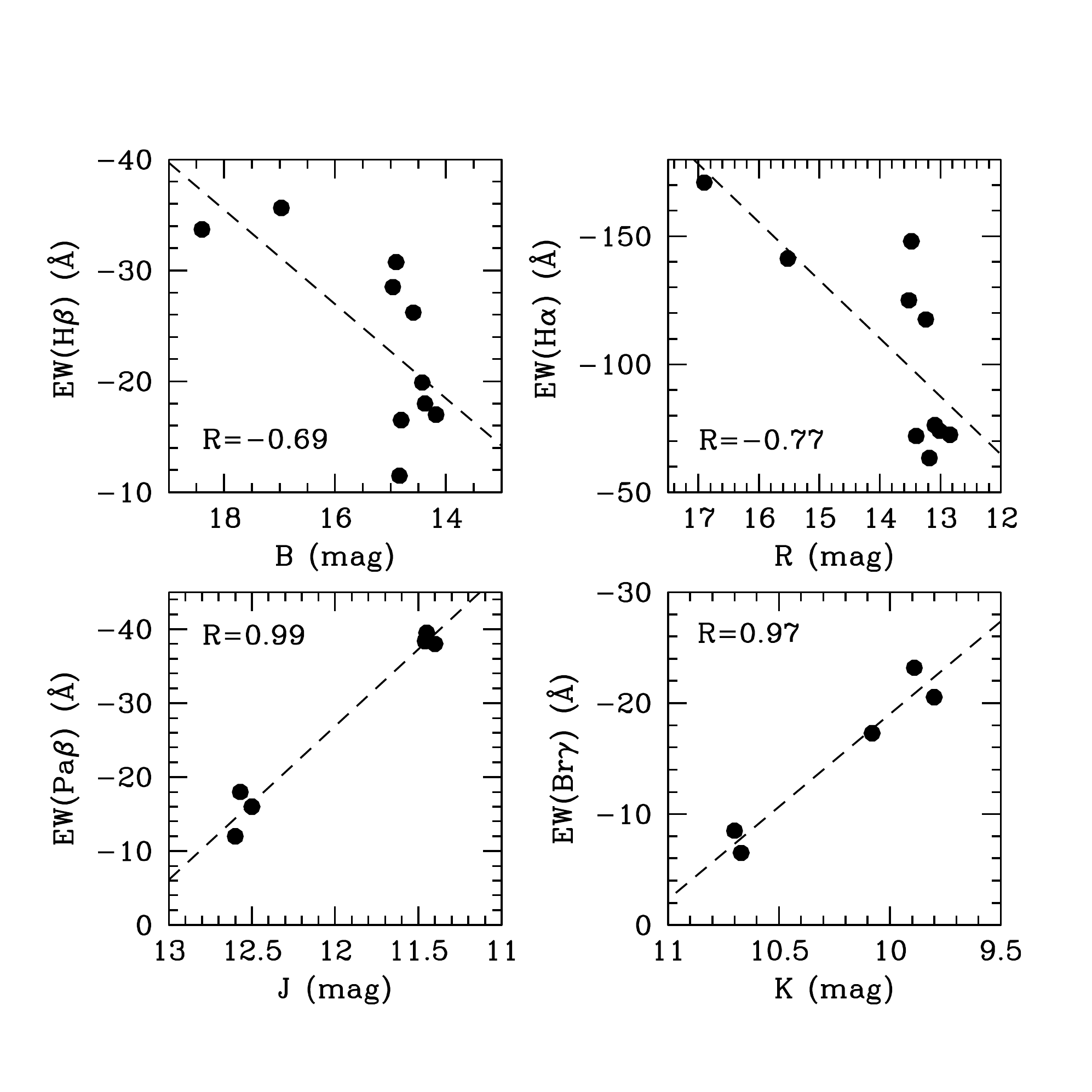}
\caption{ \label{fig:fig5} EW of optical (upper panels) and near-infrared (lower panels) \hi\, recombination lines as a function of the 
underlying continuum (in magnitudes). In each panel, the linear fit through the data is shown with a dashed line, and the correspondent regression coefficient is indicated.}
\end{figure}

\subsubsection{EW vs. continua}\label{sec:sec3.2.1}     

As a benefit of dealing with outbursts of the same source, we can examine in detail the simultaneous  variability of the continuum and line emission, removing any significant contamination that can arise when considering multiple sources (for example having different disk inclination).  
In particular, we consider the line emission that likely originates in the accretion columns close to the stellar surface (prominent \hi\, recombination, Muzerolle et al. 1998, Alcal{\'a} et al. 2014, 2017) in phases characterized by a different level of activity. As a premise, a correlation between H$\alpha$ and H$\beta$ fluxes and $R$ and $B$-band continua was highlighted in a sample of EXors (Lorenzetti et al. 2015 and references therein). This supports the idea that accretion-driven mechanisms explain both line and continuum variability.

Here, we compare the Equivalent Width (EW) of \hi\, recombination lines with the underlying continuum.
Having examined all the spectra of V1118 Ori that we have collected so far (Lorenzetti et al. 2006, 2015, Giannini et al. 2016, 2017), we show in Figure~\ref{fig:fig5} the EW variations of both the optical (H$\beta$ and H$\alpha$) and near-IR (Pa$\beta$ and Br$\gamma$) lines as a function of the magnitude of the continuum, taken almost simultaneously. Confirming the results of previous studies (PV Cep, Cohen et al. 1981; V1647 Ori, Acosta-Pulido et al. 2007), there is an anti-correlation in the optical bands. More interestingly, it is the well-defined correlation between Pa$\beta$ and Br$\gamma$ fluxes and $J$ and $K$ magnitudes (regression coefficients of -0.99 and -0.97).
This dual behavior of optical and near-infrared lines is compatible with the different regions where the lines and the continuum originate. The anti-correlation in the $B$ and $R$ bands indicates that the optical continuum varies more (or faster) than the H$\beta$ and H$\alpha$ line emission. This can be naturally explained considering that the continuum emission is related to the heating of the photosphere in the accretion shock, while \hi\, lines arise in the cooling of the gas in the accretion columns. Therefore, the observed anti-correlation simply reflects different heating and cooling times.  
Conversely, $J$- and $K$-band continua originate in the innermost regions of the circumstellar disk. Their variation (although contributed by multiple processes, eg. viscous heating) primarily follows the heating (or cooling) of the disk in response to the increase (or decline) in the temperature of the photosphere (Lorenzetti et al. 2007). As a consequence, $J$- and $K$-band continua are subject to a lower variation than \hi\, lines,  being the disk more distant from the star than the accretion columns.
Anyhow, both the correlation and the anti-correlation exclude an extinction-driven origin of the variability, since in that case the EW values should be constant for any fluctuation of the continuum.

\subsubsection{Mass accretion rate and extinction variability}\label{sec:sec3.2.2}     
 %-------------------------------------- Figura 6
  \begin{figure}
   \centering
   \includegraphics[width=10cm]{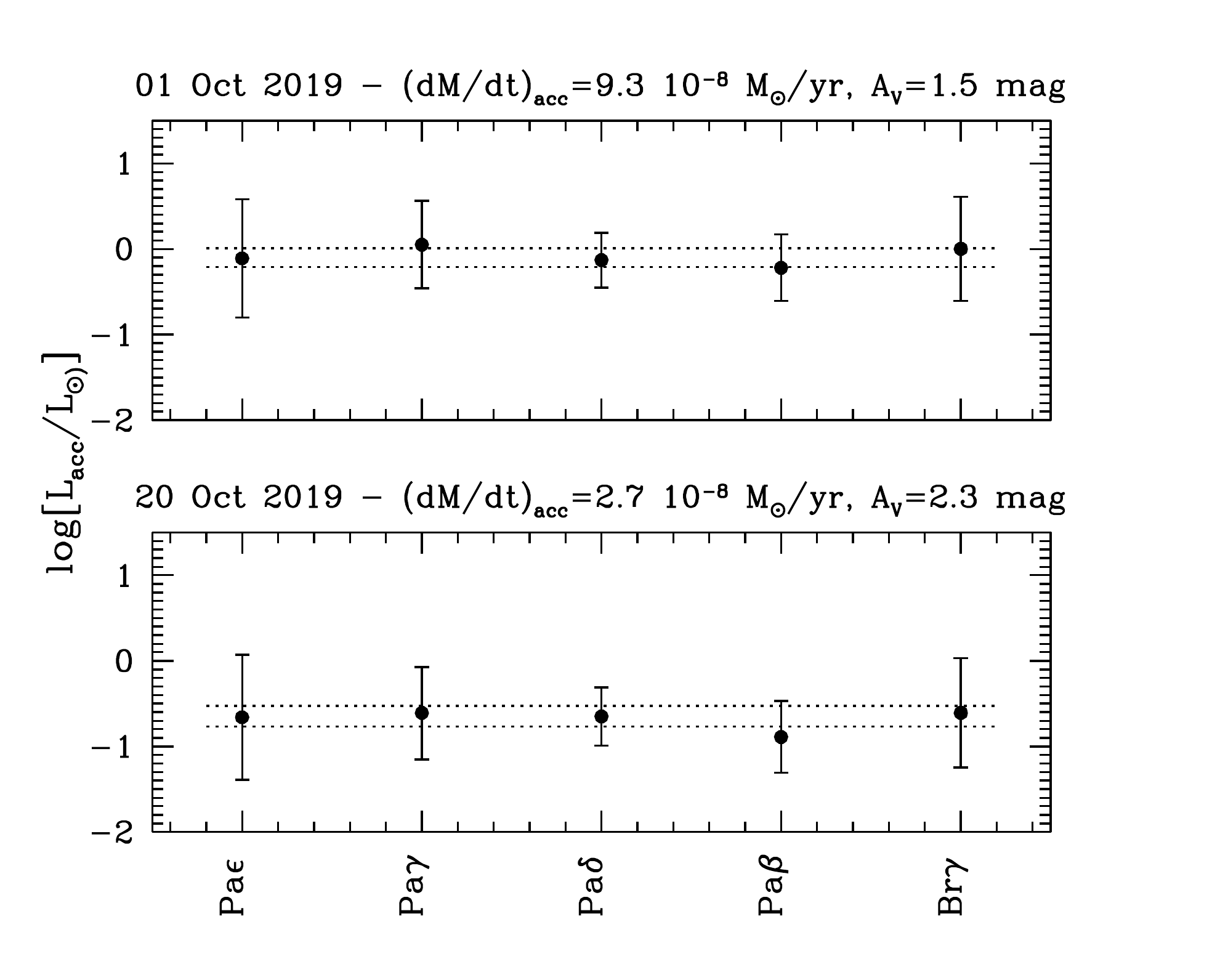}
   \caption{ \label{fig:fig6}    
   Accretion luminosity derived by applying the Alcal{\'a} et al. (2017) empirical relationships between \lacc\,  and the luminosity of the indicated \hi\, lines. \lacc\, is derived by iteratively adjusting the \av\, value to minimize the spread among the individual \lacc\,({\it i}) derived by each line {\it i}. The error bars in the data points take into account the flux errors of Table\,\ref{tab:tab3} along with the uncertainties in the Alcal{\'a} et al. relationships. The dashed lines in each panel delimit the statistical error on \lacc\,. On top of each panel is reported the date of the observation, the \macc\, derived  from \lacc\, applying the Gullbring et al. (1988) formula, and the fitted \av\,.}
 \end{figure}

The mass accretion rate on the dates of the spectroscopic observations was derived by the empirical relationships  between accretion luminosity (\lacc) and luminosity of lines  in the accretion columns ($L_{\mathrm{line}}$), determined in a sample of T Tauri stars in Lupus by Alcal{\'a} et al.\,(2017). Among the lines observed in the 2019 spectra, those for which such relationships are available are the Paschen lines from Pa$\epsilon$ to Pa$\beta$ plus the Br$\gamma$ line. First, each line flux was converted into $L_{\mathrm{line}}$ by adopting a distance to V1118 Ori of 400 pc (Muench et al. 2008). Then, \lacc\, has been estimated together with the visual extinction A$_V$ with a recursive fitting method (for a detailed explanation see Giannini et al. 2018).  The mass accretion rate, \macc\,, derives from \lacc\, through the formula of Gullbring et al. (1998), taking \mstar\,=\,0.4 \msun\, \rstar=\,1.29 \rsun\, (Hillenbrand 1997, Stassun et al. 1999) and a typical disk inner radius $R_\mathrm{in}$\,=\,5 \rsun\,.

Fit results are in Figure\,\ref{fig:fig6} and listed in Table\,\ref{tab:tab5}, where they are compared with the determinations of other outbursts and quiescence periods
we have derived by applying the same method. For the 2019 outburst, we measure a maximum  \macc\, of $\sim$ 10$^{-7}$ \msunyr\,, quite similar to the value reached in the outburst of 2015, and a factor of ten less than that of 2005. In Table\,\ref{tab:tab5} we give also the fitted values of the visual
 extinction. Interestingly, except for the last measurement in October 2019, the A$_V$ variation follows the evolution of brightness. The maximum A$_V$ of 2.5 mag is found in 2014 at the end of a long quiescence period that lasted about ten years, while A$_V$ $\sim$ 1.5 mag is
   measured at the peak of all three outbursts. The minimum A$_V$ ($<$ 1 mag) is reached during the decline phase both in
    2005 and 2016. This A$_V$ variability might be explained in the scenario in which the onset of an outburst occurs when the material accumulated in the inner disk reaches a threshold value that corresponds to the maximum of A$_V$. The outburst continues until the reservoir of material is
     over (namely the minimum of A$_V$) and then the material starts to accumulate again with a consequent increase of A$_V$. Support for this scenario might come in the case that a significant area of the V1118 Ori (circumbinary) disk  were intercepted by the line of sight. Unfortunately, the inclination  of the V1118 Ori disk remains unknown even after dedicated ALMA observations (Cieza et al. 2018).

%%%%%%%%%%%%%%Table 5 %%%%%%%%%%%%%%%%%%%%%%%%%%%%

\begin{table}
\begin{center} 
\caption{Extinction, accretion luminosity and mass accretion rate between 2005 and 2019. \label{tab:tab5}} 
\centering
\footnotesize
\begin{tabular}{ccccc}
\hline\hline%
Date              & status          & A$_V$     & \lacc            &  \macc                 \\ 
                  &                 & (mag)     & (\lsun)          &  (10$^{-7}$ \msunyr)   \\
\hline
Late 2005$^a$     & outburst peak      &  1.4 &   $-$            & $\sim$ 10     \\
11 Sep  2005      & declining          &  0.7      &   1.3            & 1.70          \\
25 Mar  2014      & quiescence         &  2.5      &   0.02           & 0.025         \\
12 Jan  2016      & outburst peak      &  1.5      &   1.1            & 1.25          \\
31 Jan  2016      & outburst peak      &  1.5      &   0.85           & 1.09          \\
04 Dec  2016      & quiescence         &  0.5      &   0.025          & 0.035         \\
01 Oct  2019      & outburst peak      &  1.5      &   0.72           & 0.93          \\
20 Oct  2019      & declining          &  2.3      &   0.21           & 0.27          \\
\hline\hline
\end{tabular}
\end{center}
\tablefoot{$^a$ Taken from Audard et al. (2005).}
\end{table}

\section{Final remarks}\label{sec:sec4}

We have presented the optical photometry and near-IR spectroscopy of the last outburst of the classical EXor V1118 Ori, which occurred in the period January-October 2019. This one is the fifth well-sampled outburst, since 1989, characterized by a a variation in brightness greater than 3 $g$-band magnitudes. Comparing the properties of the last event with those of the previous ones, we can summarize similarities and salient differences:

\begin{itemize}
\item  The outburst amplitude of the last event is similar to that of 1989, 1998 and 2015. The 2005 outburst remains the brightest. The duration is roughly the same (less than one year) for all.
\item The historical light curve shows no evident periodicity, thus favoring accretion-driven rather than extinction-driven mechanisms as triggers of the outburst.
\item The latest outburst showed different rise and decrease speed of 0.018 and 0.031 mag/day, as opposed to the 2005 outburst, when the increase was faster (by more than a factor two) compared to the decrease. This hints at different modalities that both regulate heating and cooling and vary from event to event.
\item The last three outbursts showed a bluer and bluer [$g-r$] color that we interpret with a gradual increase in the temperature of the photosphere. No significant variation is found in the [$r-i$] color.  
\item The low-resolution near-IR spectrum, taken near the peak of brightness, is similar to that of 2015 (in 2005 no near-infrared spectra were taken). Main emission lines are the \hi\, Paschen and Brackett lines. As observed in 2015, \hei\ 1.08 $\mu$m has a P Cyg-like profile that signals outflow activity. Faint metallic lines emission is present.
\item The equivalent width of the \hi\, optical lines (H$\alpha$ and H$\beta$) are anti-correlated with the $R$- and $B$-band continua, while the EW of the near-IR lines (Pa$\beta$, Br$\gamma$) are well correlated with $J$- and $K$-band continua.
This indicates all the lines (both optical and near-IR) originate at the same distance from the star (i.e. in the accretion columns), intermediate between the stellar surface and the inner disk, where most of the optical and infrared continua are emitted, respectively.
\item The evolution of the mass accretion rate of the last outburst is similar to that of 2015, with a peak of \macc\, around 10$^{-7}$ \msunyr\, and a decrease down to  $\sim$ 3 10$^{-8}$  \msunyr\, in about a month. The evolution of the extinction during the last three outbursts was also probed and tentatively interpreted in the context of the variability of the source.
\end{itemize}

\begin{acknowledgements}
The authors sincerely thank Mario Lattanzi for providing information on the Gaia observations of V1118 Ori and Gianluca Li Causi for his suggestions and constructive discussions. We aknowledge the italian LBT and REM teams for their support for both observations and data reduction. This work is  based on observations obtained with with different instruments: [1] the Samuel Oschin 48-inch Telescope at the Palomar Observatory as part of the Zwicky Transient Facility project; [2] the Large Binocular Telescope (LBT). The LBT is an international collaboration among institutions in the United States, Italy and Germany. LBT Corporation partners are: The University of Arizona on behalf of the Arizona university system; Istituto Nazionale di Astrofisica, Italy; LBT Beteiligungsgesellschaft, Germany, representing the Max-Planck Society, the Astrophysical Institute Potsdam, and Heidelberg University; The Ohio State University, and The Research Corporation, on behalf of The University of Notre Dame, University of Minnesota and University of Virginia; [3] the Rapid Eye Mount (REM) Telescope, La Silla, Chile.
\end{acknowledgements}

{}
\end{document}